\documentclass{tlp}
\usepackage{aopmath}
\usepackage{graphicx}
\usepackage{amssymb}
\usepackage{url}

\usepackage{listings}
\newtheorem{example}{Example} 

\newcommand{\eofex}{\mbox{}\nobreak\hfill\hspace{0.5em}$\blacksquare$}

\newcommand{\at}[1]{\texttt{#1}}
\newcommand{\system}[1]{\textsc{#1}}
\newcommand{\subt}[1]{\mathrm{sub}(#1)}
\newcommand{\taxa}[1]{\mathrm{tx}(#1)}
\newcommand{\qof}[1]{\mathrm{qt}(#1)}
\newcommand{\proj}[2]{#1_{#2}}

\newcommand{\Union}{\bigcup}

\newcommand{\el}[1]{\vspace{#1\baselineskip}}

\begin{document}
\bibliographystyle{acmtrans}

\long\def\comment#1{}

\title{Optimizing Phylogenetic Supertrees Using Answer Set Programming}

\author[L. Koponen et al.]
{LAURA KOPONEN and EMILIA OIKARINEN and TOMI JANHUNEN\\
HIIT and Department of Computer Science\\
Aalto University \\
P.O.\ Box 15400, FI-00076 AALTO, Finland \\
\email{\{Laura.J.Koponen, Emilia.Oikarinen,
Tomi.Janhunen\}@aalto.fi}
\and LAURA S\"AIL\"A \\
Department of Geosciences and Geography\\
University of Helsinki\\
P.O.\ Box 64, FI-00014 University of Helsinki, Finland \\
\email{Laura.Saila@helsinki.fi}
}
\maketitle

\label{firstpage}

\begin{abstract}
The supertree construction problem is about combining several
phylogenetic trees with possibly conflicting information into a single
tree that has all the leaves of the source trees as its leaves and the
relationships between the leaves are as consistent with the source
trees as possible. 
This leads to an optimization problem that is computationally
challenging and typically heuristic methods, such as matrix
representation with parsimony (MRP), are used.
In this paper we consider the use of answer set programming to solve  
the supertree construction problem in terms of two alternative encodings.
The first is based on an existing encoding of trees using 
substructures known as
quartets, while the other novel encoding captures the relationships
present in
trees through direct projections. 
We use these encodings to compute a genus-level supertree for
the family of cats (Felidae).
Furthermore, we compare our results to recent supertrees obtained  
by the MRP method. 
\end{abstract}


\section{Introduction}

In the {\em supertree construction problem}, one is given a set of
phylogenetic trees ({\em source trees}) with overlapping sets of leaf
nodes (representing {\em taxa}) and the goal is to construct a single
tree that respects the relationships in individual source trees as
much as possible \cite{bininda2004phylogenetic}.
The concept of respecting the relationships in the source trees varies
depending on the particular supertree method at hand.
If the source trees are compatible, i.e., there is no conflicting
information regarding the relationships of taxa in the source trees,
then supertree construction is easy \cite{aho1981inferring}. However,
this is rarely the case. It is typical that source trees obtained 
from different studies contain conflicting information, which makes 
supertree optimization a computationally challenging problem
\cite{foulds1982steiner,day1986computational,byrka2010new}.  

One of the most widely used supertree methods is matrix
representation with parsimony (MRP)
\cite{baum1992combining,ragan1992phylogenetic} 
in which source trees are encoded into a binary matrix, 
and maximum parsimony analysis is then used to construct a tree. Other
popular methods include matrix representation with flipping
\cite{chen2003flipping} and MinCut supertrees
\cite{semple2000supertree}. 
There is some criticism towards the accuracy and performance of
MRP, indicating input tree size and shape biases 
and varying results depending on the chosen matrix representation 
\cite{purvis1995modification,wilkinson2005shape,CLA:CLA514}. 
An alternative approach is to directly consider the {\em topologies}
induced by the source trees, for instance, using {\em quartets}
\cite{piaggio2004quartet} or {\em triplets} \cite{bryant1997building},
and  try to maximize the satisfaction of these topologies resulting in
{\em maximum quartet} (resp. {\em rooted triplet}) {\em consistency
  problem}. 
The quartet-based methods have received increasing interest
over the last few years \cite{snir2012quartet} and the quality of
supertrees produced have been shown to be on a par with MRP trees
\cite{swenson2011experimental}. 

There are a number of constraint-based approaches tailored for the
{\em phylogeny reconstruction} problem 
\cite{DBLP:conf/lpar/KavanaghMTMZG06,brooks2007inferring,wu2007quartet,sridhar2008mixed,DBLP:journals/fuin/MorgadoM10}.  
In phylogeny reconstruction, one is given a set of sequences (for
instance gene data) or topologies (for instance quartets) as input and
the task is to build a phylogenetic tree that represents the
evolutionary history of the species represented by the input.
In \cite{brooks2007inferring}, answer set programming (ASP)
is used to find cladistics-based phylogenies, and in 
\cite{DBLP:conf/lpar/KavanaghMTMZG06,sridhar2008mixed} maximum
parsimony criteria are applied, using ASP and mixed integer
programming (MIP), respectively.  
The most closely related approach to our work 
is the one in \cite{wu2007quartet} where an ASP encoding  
for solving the maximum quartet consistency problem for
phylogeny reconstruction is presented.
The difference to supertree optimization is that in phylogeny
reconstruction, typically almost all possible quartets over all sets
of four taxa are available, with possibly some errors. 
In supertree optimization the overlap of source trees is limited and
the number of quartets obtained from source trees is much smaller than
the number of possible quartets for the supertree.
For example, the supertree shown in Figure~\ref{fig:supertrees}
(right), with 34 leaf nodes, displays 46~038 different quartets, while
the source trees used to construct it only contributed 11~319 distinct
quartets, some of which were mutually incompatible.
In \cite{DBLP:journals/fuin/MorgadoM10} a constraint programming
solution is introduced for the maximum quartet consistency problem.
There are also related studies of supertree
optimization based on constraint reasoning.
In \cite{DBLP:conf/bcb/ChimaniRB10} a MIP solution for
{\em minimum flip supertrees} is presented, and
in \cite{DBLP:conf/cp/GentPSW03} constraint programming is used to 
produce {\em min-ultrametric trees} using triplets.
However, in both cases the underlying problem is polynomially
solvable.
Furthermore, ASP has also been used to formalize phylogeny-related
queries in \cite{DBLP:conf/iclp/LeNPS12}.

In this paper we solve the supertree optimization problem in terms of
two alternative ASP encodings.  
The first encoding is based on quartets and is similar to the one in
\cite{wu2007quartet}, though instead of using an ultrametric matrix, 
we use a direct encoding to obtain the tree topology. However, the
performance of the quartet-based encoding does not scale up. 
Our second encoding uses a novel approach capturing the relationships
present in trees through projections, formalized in terms of the {\em
maximum projection consistency problem}. 
We use these encodings to compute a genus-level supertree for the
family of cats (Felidae) and compare our results to
recent supertrees obtained from the MRP method.

The rest of this paper is organized as follows. 
We present the supertree problem in Section
\ref{sect:bg}, and
introduce our encodings for supertree optimization in Section
\ref{sect:enc}. 
In Section \ref{sect:exp}, we first compare the efficiency of the encodings,
and then use the projection-based encoding to compute a genus-level
supertree for the family of cats (Felidae).
We compare our supertrees to recent supertrees obtained using the MRP
method.
Finally, we present our conclusions in Section \ref{sect:conc}. 

\section{Supertree problem} 
\label{sect:bg}

A \emph{phylogenetic tree} of $n$ taxa has
exactly $n$ leaf nodes, each corresponding to one taxon. The tree may
be \emph{rooted} or \emph{unrooted}. In this work we consider
rooted trees and assume that the root has a special taxon called
{\em outgroup} as its child. 
An inner node is \emph{resolved} if it has exactly two
children, otherwise it is  \emph{unresolved}.
If a tree contains any unresolved nodes, it is unresolved; otherwise,
it is resolved.
\emph{Resolution} is the ratio of resolved inner nodes in a phylogenetic
tree.
A higher resolution is preferred, as this means that more is known about
the relationships of the taxa.  

\begin{figure}
\centering
\begin{minipage}{.4\textwidth}
  \centering
  \includegraphics[width=.8\textwidth]{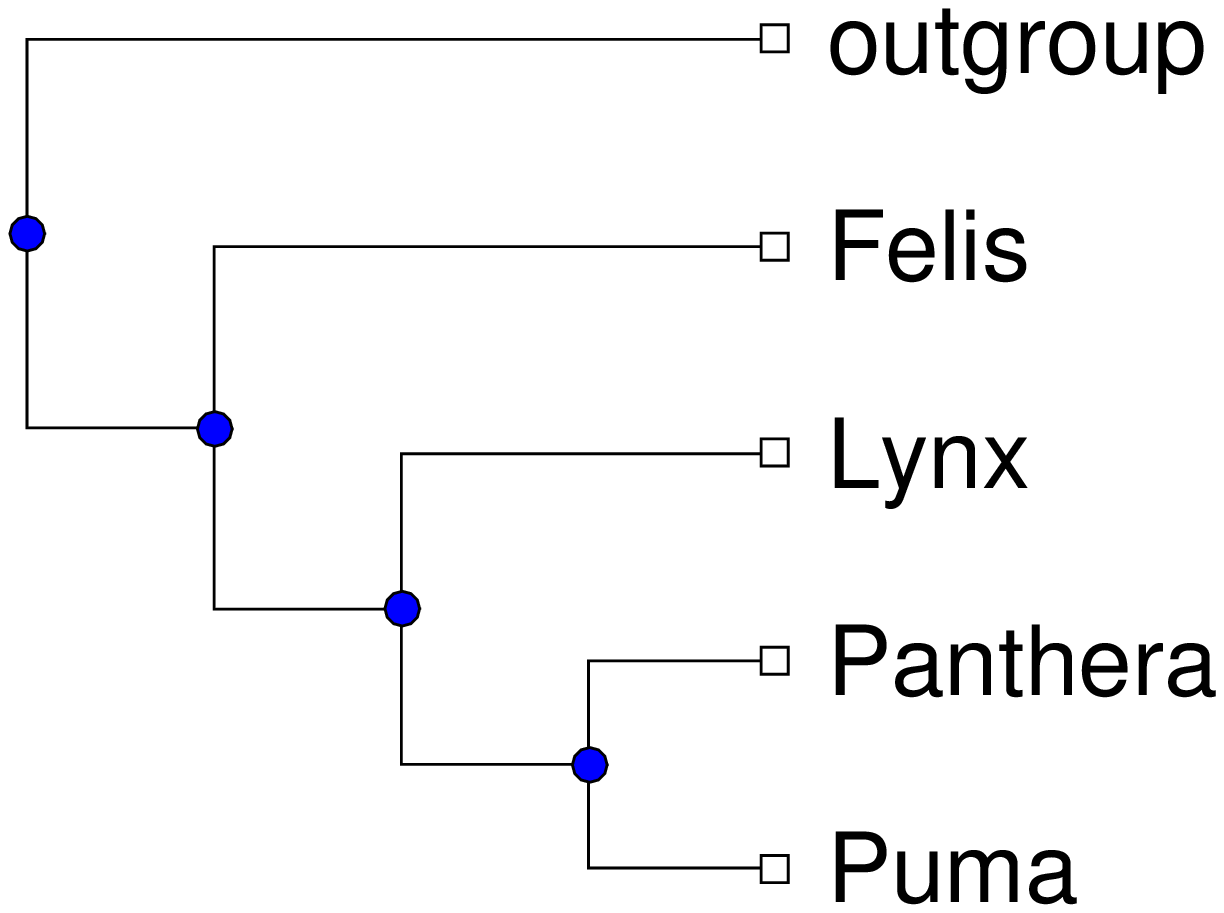}
\end{minipage}%
\begin{minipage}{.4\textwidth}
  \centering
  \includegraphics[width=.8\textwidth]{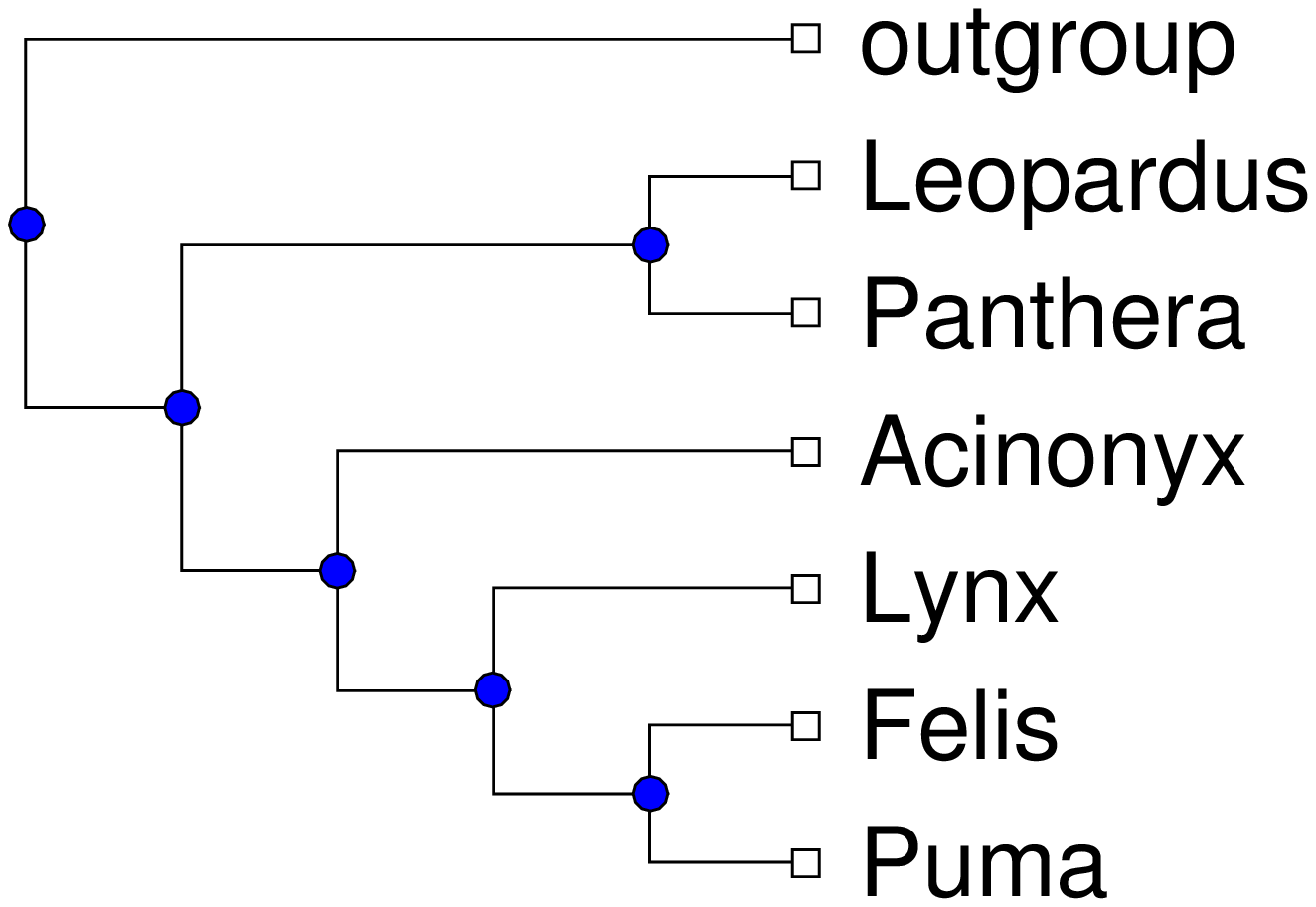}
\end{minipage}
\caption{Phylogenetic trees from \protect\cite{fulton2006molecular}
  on the left and from \protect\cite{flynn2005molecular} on the
  right, abstracted to genus-level (for more details, see Section
  \ref{sect:exp}).} 
\label{fig:trees}
\end{figure}

The problem of combining a set of phylogenetic trees with (partially)
overlapping sets of taxa into a single tree is known as the
\emph{supertree construction problem}.  
In the special case where each source tree contains exactly the same
set of species, it is also called the \emph{consensus
  tree problem} \cite{steel2000simple}. 
In order to combine trees with different taxa, one needs a way to split
the source trees into smaller structures which describe the
relationships in the trees at the same time.
There are several ways to achieve this, for instance by using triplets
(rooted substructures with three leaf nodes) or quartets (unrooted
substructures with four leaf nodes).

A {\em quartet (topology)} is an unrooted topological substructure of
a tree. The quartet $((I, J),\allowbreak (K, L))$ is in its
\emph{canonical representation} if $I < J$, $I < K$, and $K < L$,
where ``$<$'' refers to the alphabetical ordering of the names of the
taxa. From now on, we will consider canonical representations of
quartets. 
We say that a tree $T$ displays a quartet $((I, J),\allowbreak (K,
L))$, if there is an edge in the tree $T$ that separates $T$ into two
subtrees so that one subtree contains the pair $I$ and $J$ as its
leaves and the other subtree contains the pair $K$ and $L$ as its leaves.
For any set of four taxa appearing in a resolved phylogenetic tree
$T$, there is exactly one quartet displayed by $T$.
Furthermore, we say that two phylogenetic trees $T$ and $T'$ are not
compatible, if there is a set of four taxa for which $T$ and $T'$
display a different quartet. 

\begin{example}
Consider the two phylogenetic trees in Figure \ref{fig:trees}. It is easy
to see that these trees are not compatible. For the taxa {\em Felis, Lynx,
Panthera,} and {\em Puma}, the tree on the left displays the quartet {\em
((Felis,Lynx),(Panthera,Puma))}, while the tree on the right displays the
quartet {\em ((Felis,Puma),(Lynx,Panthera))}. \eofex
\end{example}

Let $\taxa{T}$ denote the set of taxa in the leaves of a tree $T$ and 
$\qof{T}$ the set of all quartets that are displayed by $T$.
For a collection $S$ of phylogenetic trees, we define $\qof{S}$
as the multiset%
\footnote{We use multisets in order to give more weight to structures
appearing in several source trees.}
$\Union_{T\in S}\qof{T}$ and $\taxa{S}=\Union_{T\in S}\taxa{T}$. 
Given any phylogenetic tree $T$, the set $\qof{T}$ uniquely determines it
\cite{DBLP:dblp_journals/rsa/ErdsSSW99}. 

The \emph{quartet compatibility problem} is about finding out whether a set
of quartet topologies $\qof{S}$ for a collection of phylogenetic trees
$S$ is compatible, i.e., if there is a phylogeny $T$ on the taxa in
$\taxa{S}$ that displays all the quartet topologies in $\qof{S}$.
The \emph{maximum quartet consistency problem} for a supertree takes
as input a set of quartet topologies $\qof{S}$ for a collection of
phylogenetic trees $S$, and the goal is to find a phylogeny $T$ on the
taxa $\taxa{S}$ that displays the maximum number of quartet topologies
in $\qof{S}$ \cite{piaggio2004quartet}.

The topology of a tree $T$ can be captured more directly using
projections of $T$. Given a set $S\subseteq\taxa{T}$, the
\emph{projection} of $T$ \emph{with respect to} $S$, denoted by
$\proj{T}{S}$, is obtained from $T$ by removing all structure related
to the taxa in $\taxa{T}\setminus S$. This may imply that entire
subtrees are removed and non-branching nodes are deleted. We say that
$T$ \emph{displays} another tree $T'$ if
$\taxa{T'}\subseteq\taxa{T}$ and $\proj{T}{\taxa{T'}}=T'$.

\begin{example}
If the left tree in Figure \ref{fig:trees} is projected with
respect to $\{\emph{Puma},\emph{Lynx},
\emph{Felis}\}$, the following tree results: \emph{((Puma,Lynx),Felis)}.
The right
tree yields a different projection \emph{((Puma,Felis),Lynx)}
illustrating the topological difference of the trees.
\eofex
\end{example}

When comparing a phylogeny $T$ with other phylogenies, an obvious
question is which projections should be used. Rather than using arbitrary
sets $S\subseteq\taxa{T}$ for projections $\proj{T}{S}$, we suggest to
use the subtrees of $T$. We denote this set by $\subt{T}$.
It is clear that $T$ displays $T'$ for every $T'\in\subt{T}$.
Moreover, if $T$ displays $T''$ for every $T''\in\subt{T'}$ and
$\taxa{T}=\taxa{T'}$, then $T=T'$. More generally, the more subtrees
of $T'$ are displayed by $T$, the more alike $T$ and $T'$ are as
trees.
This observation suggests defining the
\emph{maximum projection consistency problem}
for a supertree in analogy to the maximum quartet consistency
problem.
The input for this problem consists of the multiset
$\subt{T_1}\union\ldots\union\subt{T_n}$
induced by a given collection $T_1,\ldots,T_n$ of phylogenetic trees.
The goal is to find a supertree $T$ such that
$\taxa{T}=\taxa{\{T_1,\ldots, T_n\}}$
and $T$ displays as many subtrees from the input as
possible---\emph{disregarding orientation}. This objective is aligned
with the quartet-based approach: if $T$ displays a particular subtree
$T'$, then it also displays $\qof{T'}$.

\begin{example}
Consider again the trees in Figure \ref{fig:trees}. 
The non-trivial subtrees of the left tree are:
\vspace{-1ex}
\begin{center}
\emph{(outgroup,(Felis,(Lynx,(Panthera,Puma)))),
      (Felis,(Lynx,(Panthera,Puma))),
      (Lynx,(Panthera,Puma)), (Panthera,Puma)}
\end{center}\vspace{-1ex}
The right tree displays
only the subtree \emph{(Panthera,Puma)} as its projection.
\eofex
\end{example}

\section{Encodings for supertree optimization}
\label{sect:enc}

We assume that the reader is familiar with basic ASP terminology and
definitions, and we refer the reader to
\cite{Baral:2003:KRR:582493,DBLP:series/synthesis/2012Gebser}
for details.
Our encodings are based on 
the input language of the
\system{gringo} 3.0.4 grounder \cite{DBLP:conf/lpnmr/GebserKOST09}
used to instantiate logic programs.
In this section, two alternative encodings for the supertree
construction problem are presented. Both encodings rely on the same
formalization of the underlying tree structure, but have
different objective functions as well as different representations for
the input data. We begin by developing a canonical representation for
phylogenies based on ordered trees in Section \ref{sect:canonical}.
The first encoding based on \emph{quartet} information is then
presented in Section \ref{sect:quartets}. The second one exploiting
\emph{projections} of trees is developed in Section \ref{sect:projections}.


\subsection{Canonical phylogenies}
\label{sect:canonical}

Our encodings formalize phylogenies as ordered trees whose leaf nodes
correspond to taxa (species or genera) of interest.
The simplest possible (atomic) tree consists of a single node.
Thus we call the leaves of the tree \emph{atoms} and formalize them in
terms of the predicate \at{atom}/1.
We assume that the number of atoms is available through the predicate
\at{atomcnt}/1, and furthermore that atoms have been ordered alphabetically
so that the first atom is accessible through the predicate \at{fstatom}/1,
while the predicate \at{nxtatom}/2 provides the successor of an atom.
These predicates can be straightforwardly expressed in the input
language of \system{gringo} and we skip their actual definitions. Full
encodings are published with tools (see Section 4).

To formalize the structure of an ordered tree with $N$ leaves, we
index the leaf nodes using numbers from 1 to $N$. Any subsequent
numbers up to $2N-1$ will be assigned to inner nodes as formalized by
lines \ref{line:node}--\ref{line:inner} of Listing \ref{code:tree}.
Depending on the topology of the tree, the number of
inner nodes can vary from $1$ to $N-1$.
In the former case, the tree has an edge from the root to every leaf but a
full binary tree results in the latter case. If viewed as phylogenies, the
former leaves all relationships unresolved whereas the latter gives a fully
resolved phylogeny.
\begin{lstlisting}[label=code:tree,frame=single,float=t,%
caption={An ASP Encoding of Directed Trees/Forests\el{0.5}},escapechar=|]
% Domains
node(1..2*N-1) :- atomcnt(N).|\label{line:node}|
leaf(X) :- node(X), X<=N, atomcnt(N).|\label{line:leaf}|
inner(X) :- node(X), X>N, atomcnt(N).|\label{line:inner}|
pair(X,Y) :- inner(X), node(Y), X>Y.|\label{line:pair}|

% Choose edges
{ edge(X,Y): pair(X,Y) } 2*N-2 :- atomcnt(N).|\label{line:choice}|
:- edge(X,Z), edge(Y,Z), pair(X;Y,Z), X<Y.|\label{line:nodag}|
:- edge(X,Y), pair(X,Y), inner(Y), not edge(Y,Z): pair(Y,Z).|\label{line:dead}|

% Assign atoms to leaves
asgn(1,A) :- node(1), fstatom(A).|\label{line:first-leaf}|
asgn(N+1,B) :- node(N), asgn(N,A), nxtatom(A,B).|\label{line:next-leaf}|
\end{lstlisting}
The predicate \at{pair}/2 defined in line \ref{line:pair} declares
that the potential edges of the tree always proceed in the descending
order of node numbers. This scheme makes loops impossible and
prohibits edges starting from leaf nodes. The rule in line
\ref{line:choice} chooses at most $2N-2$ edges for the tree up to
$2N-1$ nodes. The constraint in line \ref{line:nodag} ensures that a
directed tree/forest rather than a directed acyclic graph is obtained.
The purpose of the constraint in line \ref{line:dead} is to deny
branches ending at inner nodes.
The fixed assignment of atoms to leaf nodes $1\ldots N$
according to their alphabetical order takes place in lines
\ref{line:first-leaf}--\ref{line:next-leaf}
using predicates \at{fstatom}/1 and \at{nxtatom}/2.
This is justified by a symmetry reduction, since $N!$
different assignments to leaf nodes would be considered otherwise
and no tree topology is essentially ruled out. 

However, as regards tree topologies themselves, further symmetry
reductions are desirable because the number of optimal phylogenies
can increase substantially otherwise.
Listing \ref{code:canonical} provides conditions for a canonical
ordering for the inner nodes. The \at{order}/2 predicate defined in
lines \ref{line:o1}--\ref{line:o2} captures pairs of inner nodes that
must be topologically ordered in a tree being constructed. The
\at{ireach}/2 predicate defined by rules in lines \ref{line:ir1} and
\ref{line:ir2} gives the \emph{irreflexive} reachability relation for
nodes, i.e., a node is not considered reachable from itself. The
constraint in line~\ref{line:order} effectively states that the
numbering of inner nodes must follow the depth-first descending order,
i.e., any inner nodes \at{X} below \at{Y} must have higher numbers
than \at{Z}.
The remaining degree of freedom concerns the placement of leaves to
subtrees. To address this, we need to find out the \emph{minimum}%
\footnote{Recall that the numbering of leaf nodes corresponds to the
alphabetical ordering of the taxa.}
leaf (node) for each subtree. The \at{min}/2 predicate defined in
lines \ref{line:min1}--\ref{line:min2} captures the actual minimum
leaf \at{Y} beneath an inner node \at{X}.  The orientation constraint
in line~\ref{line:orient} concerns inner nodes \at{Y} and \at{Z}
subject to topological ordering, identifies the minimum leaf \at{W} in
the subtree rooted at \at{Z}, and ensures that this leaf is smaller
than any leaf \at{V} in the subtree rooted at~\at{Y}. This also covers
the case that \at{V} is the respective minimum leaf under \at{Y}. The
orientation constraint above generalizes that of
\cite{brooks2007inferring}
for non-binary trees and we expect that canonical trees will have
further applications beyond this work.

\begin{lstlisting}[label=code:canonical,frame=single,float=t,%
caption={Encoding for Canonical Phylogenies\el{0.5}},escapechar=|]
% Depth-first ordering on internal nodes
order(Y,Z) :- edge(X,Y), edge(X,Z), pair(X,Y;Z), inner(Y;Z), |\label{line:o1}|
              Y>Z, not edge(X,W): Y>W: W>Z: pair(X,W). |\label{line:o2}|
ireach(X,Y) :- edge(X,Y), pair(X,Y).|\label{line:ir1}|
ireach(X,Y) :- ireach(X,Z), edge(Z,Y), pair(Z,Y).|\label{line:ir2}|
:- order(Y,Z), pair(Y,Z), ireach(Y,X), inner(X), X<Y.|\label{line:order}|

% Determine the orientation of leaf nodes
min(X,Y) :- ireach(X,Y), inner(X), leaf(Y),|\label{line:min1}|
            not ireach(X,Z): Z<Y: leaf(Z).|\label{line:min2}|
:- order(Y,Z), pair(Y,Z), ireach(Y,V), min(Z,W), leaf(V;W), V<W.|\label{line:orient}|

% Constraints for phylogenies
:- unused(X), used(Y), inner(X;Y), X<Y.|\label{line:unused}|
:- root(X), root(Y), inner(X;Y), X<Y.|\label{line:double-root}|
:- not root(X): inner(X).|\label{line:inner-root}|
:- leaf(X), not edge(Y,X): pair(Y,X).|\label{line:connect-leaves}|
:- inner(X), root(X), not outgroup(X).|\label{line:og1}|
:- inner(X), not root(X), outgroup(X).|\label{line:og2}|
:- edge(X,Y), pair(X,Y), not edge(X,Z): pair(X,Z): Z!=Y.|\label{line:nounary}|
\end{lstlisting}

Finally, there are some further requirements specific to phylogenies.
We assume that certain subsidiary predicates have already been
defined. The predicate \at{root}/1 is used to identify root nodes.
Inner nodes that remain completely disconnected are marked as unused by the
predicate \at{unused}/1.
Otherwise, the node is in use as captured by \at{used}/1.
Moreover, a node is an \emph{outgroup} node, formalized by
\at{outgroup}/1, if it is assigned to the special \emph{outgroup} taxon or
one of its child nodes is so assigned (cf.~Figure~\ref{fig:trees}).
Lines~\ref{line:unused}--\ref{line:nounary} list the additional constraints
for a phylogeny.
Only the highest numbers are allowed for unused nodes (line
\ref{line:unused}). The root must be a unique inner node (lines
\ref{line:double-root} and \ref{line:inner-root}). Every leaf must be
connected (line \ref{line:connect-leaves}). The special outgroup leaf must
be associated with the root node (lines \ref{line:og1} and \ref{line:og2}).
Every inner node that is actually used must have at least two children
(line \ref{line:nounary}): the denial of \emph{unary} nodes is justified
because they are not meaningful for phylogenies.


\subsection{Quartet-based approach}
\label{sect:quartets}

The first encoding is \emph{quartet-based}. Each source tree is represented
as the set of all quartets that it displays. The predicate \at{quartet}/4
represents one input quartet in canonical form.
Listing~\ref{code:quartets} shows the objective function for the
quartet encoding. For each quartet appearing in the input, we check if
it is satisfied by the current output tree candidate.
The auxiliary predicate \at{reach}/2 marks reachability from inner nodes to
atoms (species) assigned to leaves.
The output tree is rooted, so given any inner node \at{X} in the tree,
there is a uniquely defined subtree rooted at \at{X}, and
\at{reach(X,A)} is true for any atom \at{A} corresponding to a leaf
node of the subtree.
A quartet consisting of two pairs is satisfied by the output tree, if
for one pair there exists at least one inner node \at{X} such that the
members of the pair are descendants of \at{X}, while the members of
the other pair do not appear in that subtree.

The predicate \at{quartetwt}/5 assigns a weight to each quartet structure.
In the unweighted case, this weight is equal to the number of source trees
that display the quartet. In the weighted case, source trees stemming from
computational studies based on molecular input data were weighted up by a
factor of four. For example, if a particular quartet was present in
three source trees, two of which were from molecular studies while the
third one was not, the total weight would be $4 + 4 + 1$.

\begin{lstlisting}[label=code:quartets,frame=single,float=t,%
caption={Optimization function for the quartet
encoding\el{0.5}},escapechar=|]
reach(X,A) :- inner(X), ireach(X,Y), asgn(Y,A), atom(A).

% Maximize number of satisfied quartets
satisfied(A1,A2,A3,A4) :- quartet(A1,A2,A3,A4), inner(X),
   reach(X,A1), reach(X,A2), not reach(X,A3), not reach(X,A4).
satisfied(A1,A2,A3,A4) :- quartet(A1,A2,A3,A4), inner(X),
   reach(X,A3), reach(X,A4), not reach(X,A1), not reach(X,A2).

#maximize [ satisfied(A1,A2,A3,A4)=W: quartetwt(A1,A2,A3,A4,W) ].
\end{lstlisting}


\subsection{Projection-based approach}
\label{sect:projections}

The second encoding is based on direct \emph{projections} of trees and
the idea is to identify which inner nodes in the selected phylogeny
correspond to subtrees present in the input trees. Input trees are
represented using a function symbol \at{t} as a tree constructor.
For instance, the leftmost tree in Figure \ref{fig:trees} is
represented by a term
\begin{equation}
\label{eq:phylogeny-as-term}
\at{t(outgroup,t(felis,t(lynx,t(panthera,puma))))}.
\end{equation}
For simplicity, it is assumed here that \at{t} always takes two arguments
although in practice, some of the input trees are non-binary, and a more
general list representation is used instead.
In the encoding, projections of interest are declared in terms of the
predicate \at{proj}/1. The predicate \at{comp}/1, defined in line
\ref{line:comp} of Listing~\ref{code:projections}, identifies
\emph{compound trees}
as those having at least one instance of the constructor \at{t}. The
set of projections is made downward closed by the rule in
line~\ref{line:proj}. For instance, \at{outgroup} and
\at{t(felis,t(lynx,t(panthera,puma)))} are projections derived from
(\ref{eq:phylogeny-as-term}) by a single application of this rule.  In
line~\ref{line:atom}, \emph{atoms} are recognized as trivial tree
projections with no occurrences of \at{t} such as \at{outgroup}
above.

\begin{lstlisting}[label=code:projections,frame=single,float=t,%
caption={Projection-Based Optimization of the Phylogeny\el{0.5}},escapechar=|]
% Projections of the phylogeny
comp(t(T1,T2)) :- proj(t(T1,T2)).|\label{line:comp}|
proj(T1;T2) :- comp(t(T1,T2)).|\label{line:proj}|
atom(X) :- proj(X), not comp(X).|\label{line:atom}|

% Reachability from a node to a projection
reach(X,T) :- node(X), asgn(X,T), proj(T).|\label{line:reach1}|
reach(X,T) :- ireach(X,Y), node(X;Y), reach(Y,T), proj(T).|\label{line:reach2}|

% Assign compound trees to inner nodes
asgn(X,T) :- inner(X), used(X), not denied(X,T), comp(T).|\label{line:default}|
denied(X,T) :- edge(X,Y), pair(X,Y), comp(T), reach(Y,T).|\label{line:below}|
denied(X,t(T1,T2)) :- edge(X,Y), pair(X,Y), comp(t(T1,T2)),|\label{line:join1}|
                      T1<T2, reach(Y,T1), reach(Y,T2).|\label{line:join2}|
denied(X,t(T1,T2)) :- inner(X), used(X), comp(t(T1,T2)),|\label{line:t1}|
                      not reachvia(X,Z,T1): pair(X,Z).
denied(X,t(T1,T2)) :- inner(X), used(X), comp(t(T1,T2)),
                      not reachvia(X,Z,T2): pair(X,Z).|\label{line:t2}|
reachvia(X,Y,T) :- edge(X,Y), pair(X,Y), reach(Y,T), proj(T).|\label{line:via}|
:- inner(X), used(X), not asgn(X,T): comp(T).|\label{line:empty}|

% Optimize the assignment of compound trees
unassigned(T) :- comp(T), not asgn(X,T): node(X).
next(X,T) :- edge(X,Y), pair(X,Y), asgn(Y,T), proj(T).|\label{line:next}|
separated(t(T1,T2)) :- edge(X,Y), pair(X,Y), asgn(X,t(T1,T2)),|\label{line:separated1}|
                       not next(X,T1).
separated(t(T1,T2)) :- edge(X,Y), pair(X,Y), asgn(X,t(T1,T2)),
                       not next(X,T2).|\label{line:separated2}|
#minimize [ unassigned(T)=AC*W: acnt(T,AC): projwt(T,W): comp(T),|\label{line:opt2}|
	    separated(T)=W: projwt(T,W): comp(T) ]. |\label{line:opt1}|
\end{lstlisting}

The \at{reach}/2 predicate, defined in lines
\ref{line:reach1} and \ref{line:reach2}
of Listing \ref{code:projections}, generalizes the respective
predicate from Listing \ref{code:quartets} for arbitrary projections
\at{T} and includes a new base case for immediate assignments
(line \ref{line:reach1}).
A compound tree \at{T} is assigned to an inner node \at{X}
\emph{by default} (line \ref{line:default})
and the predicate \at{denied}/2 is used to specify \emph{exceptions}
in this respect.  It is important to note that if \at{edge(X,Y)} is
true, then \at{X} is an inner node and \at{used(X)} is true, too.
The first exception (line \ref{line:below}) is that \at{T} is already
assigned below \at{X} in the phylogeny. The second case (lines
\ref{line:join1}--\ref{line:join2}) avoids mapping distinct subtrees
of \at{t(T1,T2)} on the same subtree in the phylogeny. Thirdly, if
\at{t(T1,T2)} is to be assigned at inner node \at{X}, then \at{T1} and
\at{T2} must have been assigned beneath \at{X} in the phylogeny (lines
\ref{line:t1}--\ref{line:t2}). Finally, the constraint in line
\ref{line:empty} insists that each inner node is assigned at least one
projection because the node could be removed from the phylogeny
otherwise. The net effect of the constraints introduced so far is that
if \at{T1} and \at{T2} have been assigned to nodes \at{X} and \at{Y},
respectively, then \at{t(T1,T2)} is assigned to the
\emph{least common ancestor} of \at{X} and \at{Y}.

The rest of Listing \ref{code:projections} concerns the objective
function we propose for phylogeny optimization.  The predicate
\at{unassigned}/1 captures compound trees \at{T} which could not be
assigned to any inner node by the rules above. This is highly likely
if mutually inconsistent projections are provided as input. It is also
possible that a compound projection \at{t(T1,T2)} is assigned further
away from the subtrees \at{T1} and \at{T2}, i.e., they are not placed
next to \at{t(T1,T2)}. The predicate \at{separated}/1 holds for
\at{t(T1,T2)} in this case (lines
\ref{line:next}--\ref{line:separated2}).
The purpose of the objective function (line \ref{line:opt1}) is to
minimize penalties resulting from these aspects of assignments. For
unassigned compound trees \at{T}, this is calculated as the product of
the number of atoms in \at{T} and the weight%
\footnote{As before, the weight is $4$ for projections originating from
molecular studies and $1$ otherwise.}
of \at{T}. These numbers are accessible via auxiliary predicates
\at{acnt}/2 and \at{projwt}/2 in the encoding.  Separated compound
trees are further penalized by their weight (line
\ref{line:opt2}). Since the rules in lines
\ref{line:comp}--\ref{line:proj},
\ref{line:join1}--\ref{line:t2},
\ref{line:separated1}--\ref{line:separated2}
only cover binary trees they would have to be generalized for any
fixed arity which is not feasible. To avoid repeating the rules for
different arities, we represent trees as lists (of lists) in
practice.

\section{Experiments}
\label{sect:exp}

\paragraph{Data.}
We use a collection of 38 phylogenetic trees from
\cite{saila2011,saila2012} covering 105 species  
of Felidae as our source trees.\footnote{Source trees in Newick format
  are provided in the online appendix (Appendix D).} 
There are both resolved and unresolved trees, all rooted with 
{\em outgroup}, in the collection and the
number of species varies from 4 to 52.
The total number of species in the source trees makes supertree
analysis even with heuristic methods 
challenging, and 
computing the full supertree for all species at once is not feasible
with our encodings.
Thus, we consider the following simplifications of the data.
In Section~\ref{sect:species-trees} we use
{\em genus-specific projections of source trees} to compare
the efficiency of our two encodings.  
In Section \ref{sect:genus-trees} 
we reduce the size of the instance 
by considering the {\em genus-level supertree} 
as a first step towards solving the supertree problem for the Felidae
data. 

\paragraph{Experimental setting.}
We used two identical 2.7-GHz CPUs with 256~GB of RAM to compute
optimal answer sets for programs grounded by \system{gringo}~3.0.4.
The state-of-the-art solver%
\footnote{\url{http://potassco.sourceforge.net}}
\system{clasp}~3.1.2
\cite{DBLP:journals/aicom/GebserKKOSS11}
was compared with a runner-up solver
\system{wasp}\footnote{\url{http://github.com/alviano/wasp.git}}
\cite{AlvianoDFLR15}
as of 2015-06-28.
Moreover, we studied the performance of MAXSAT solvers as back-ends
using translators \system{lp2acyc}~1.29 and \system{lp2sat}~1.25
\cite{gebser2014answer}, and a normalizer \system{lp2normal}~2.18
\cite{bomanson2014improving} from the \emph{asptools}%
\footnote{Subdirectories \url{download/} and  \url{encodings/}
at \url{http://research.ics.aalto.fi/software/asp/}} 
collection.
As MAXSAT solvers, we tried \system{clasp}~3.1.2 in its MAXSAT mode
(\system{clasp-s} in Table~\ref{results2}), an \system{openwbo}-based
extension%
\footnote{\url{http://sat.inesc-id.pt/open-wbo/}}
\cite{martins2014open} of \system{acycglucose} R739
(labeled \system{acyc} in Table~\ref{results2}) also available in the
asptools collection, and \system{sat4j}%
\footnote{\url{http://www.sat4j.org/}}
\cite{le2010sat4j}
dated 2013-05-25.

\subsection{Genus-specific supertrees}
\label{sect:species-trees}

To produce genus-specific source trees for a genus $G$, we project all
source trees to the species in $G$ (and the outgroup).
Genera with fewer than five species are excluded as too trivial.
Thus, the instances of Felidae data have between 6 and 11
species each, and the number of source trees varies between 2 and 22.
In order to be able to compare the performance of different solvers
for our encodings, we compute {\em one} optimum here and use a timeout
of one hour.
In Table~\ref{results2} we report the run times for the best-performing
configuration of each solver for both 
encodings.\footnote{
We exclude \system{sat4j}, which had the longest run times, from
comparison due to space limitations.} 
Moreover, the methods based on unsatisfiable cores turned out to be
ineffective in general. Hence, branch-and-bound style heuristics were
used. 

The performance of the projection encoding scales up better than that
of the quartet encoding when the complexity of the instance grows. 
Our understanding is that in the quartet encoding the search space
is more symmetric than in the projection encoding: in principle any
subset of the quartets could do and this has to be excluded in the
optimality proof. 
On the other hand, the mutual incompatibilities of projections
can help the solver to cut down the search space more effectively.

\begin{table}
\begin{minipage}{\textwidth}
\begin{tabular}{@{$\!$}l@{$\!$}c@{$\!$}cc@{$\!$}cc@{$\!$}cc@{$\!$}cc@{$\!$}c@{$\!$}} 
& & & 
\multicolumn{2}{c}{{\sc clasp}\footnote{Options: {\tt
    --config=frumpy} (proj) and {\tt --config=trendy} (qtet)}} 
& \multicolumn{2}{c}{{\sc wasp}\footnote{Options: {\tt --weakconstraints-algorithm=basic}}}
& \multicolumn{2}{c}{{\sc acyc}\footnote{Options: {\tt -algorithm=1}
    and  {\tt -incremental=3}}} 
& \multicolumn{2}{c}{{\sc clasp-s}\footnote{Options {\tt
      --config=frumpy} (proj) and {\tt  --config=tweety} (qtet)}
}
\\
Genus & Taxa & Trees 
& qtet & proj & qtet & proj& qtet& proj & qtet & proj \\ \hline\hline
Hyperailurictis    & 6  & 2  & 0.0    &  0.0 &   0.0 &   0.0 &   0.0 &   0.0 &   0.0 &  0.0 \\
Lynx               & 7  & 8  & 0.0    &  0.0 &   0.0 &   0.1 &   0.0 &   0.0 &   0.0 &  0.0 \\
Leopardus          & 8  & 6  & 0.6    &  0.1 &   1.7 &   0.2 &   1.1 &   0.4 &   0.6 &  0.1 \\
Dinofelis          & 9  & 2  & 0.1    &  0.0 &   0.0 &   0.1 &   0.1 &   0.1 &   0.0 &  0.1 \\
Homotherium        & 9  & 3  & 0.7    &  0.0 &   0.1 &   0.1 &   0.1 &   0.0 &   0.0 &  0.0 \\
Felis              & 11 & 12 & 39.6   & 21.9 & 290.8 & 120.6 & 122.7 &  59.6 &  27.7 & 20.8 \\
Panthera           & 11 & 22 & 1395.8 & 45.6 &    -- & 456.3 &    -- & 174.6 & 944.2 & 67.1 \\\hline
\end{tabular}
\vspace{-2\baselineskip}
\end{minipage}
\caption{Time (s) to find one optimum for genus-specific data using
  different solvers using quartet (qtet) and projection (proj)
  encoding (-- marks timeout).}
\label{results2}
\end{table}

\subsection{Genus-level abstraction}
\label{sect:genus-trees}

We generate 28 trees abstracted to the genus level from the 38
species-level trees. The abstraction is done by placing each genus
$G$ under the node $N$ furthest away from the root such that all
occurrences of the species of genus $G$ are in the subtree below $N$.  
Finally, redundant (unary) inner nodes are
removed from the trees. 
The trees that included fewer than four genera were excluded.
Following \cite{saila2011,saila2012}, {\em Puma pardoides} was treated
as its own genus {\em Pardoides},
and {\em Dinobastis} was excluded as an invalid taxon. 
As further preprocessing, we removed the occurrences of genera
{\em Pristifelis}, {\em Miomachairodus}, and {\em Pratifelis} 
appearing in only one source tree each.
These so-called {\em rogue taxa} have unstable placements in the
supertree, due to little information about their placements in
relation to the rest of the taxa.
The rogue taxa can be a posteriori placed in the supertree in the
position implied by their single source tree. 
After all the preprocessing steps, our genus-level source trees have
34 genera in total and the size of the trees varies from 4 to
22~genera.  

We consider the following schemes from \cite{saila2011,saila2012}:
\begin{description}
\item[All-FM-bb-wgt] Analysis with a constraint tree separating
  the representatives of Felinae and 
  Machairodontinae into subfamilies, with weight 4 given to source
  trees from molecular studies.
\item[F-Mol] Analysis using molecular studies only and extinct species 
pruned out
(leaving 20 source trees and 15 genera, which are all representatives
of Felinae). 
\end{description}
Noticeably, the first setting allows us to split the search space and
to compute the supertree for Felinae and Machairodontinae separately.
The {\em best resolved tree} in \cite{saila2011,saila2012} was obtained 
using the MRP supertree for {\bf F-Mol} abstracted to the genus level 
as a constraint tree (scheme {\bf All-F-Mol-bb-wgt}). We include the
best resolved tree by \citeANP{saila2011}~to the comparison as well.

\begin{figure}
\centering
\begin{minipage}{.5\textwidth}
	\centering
	\includegraphics[width=\textwidth]{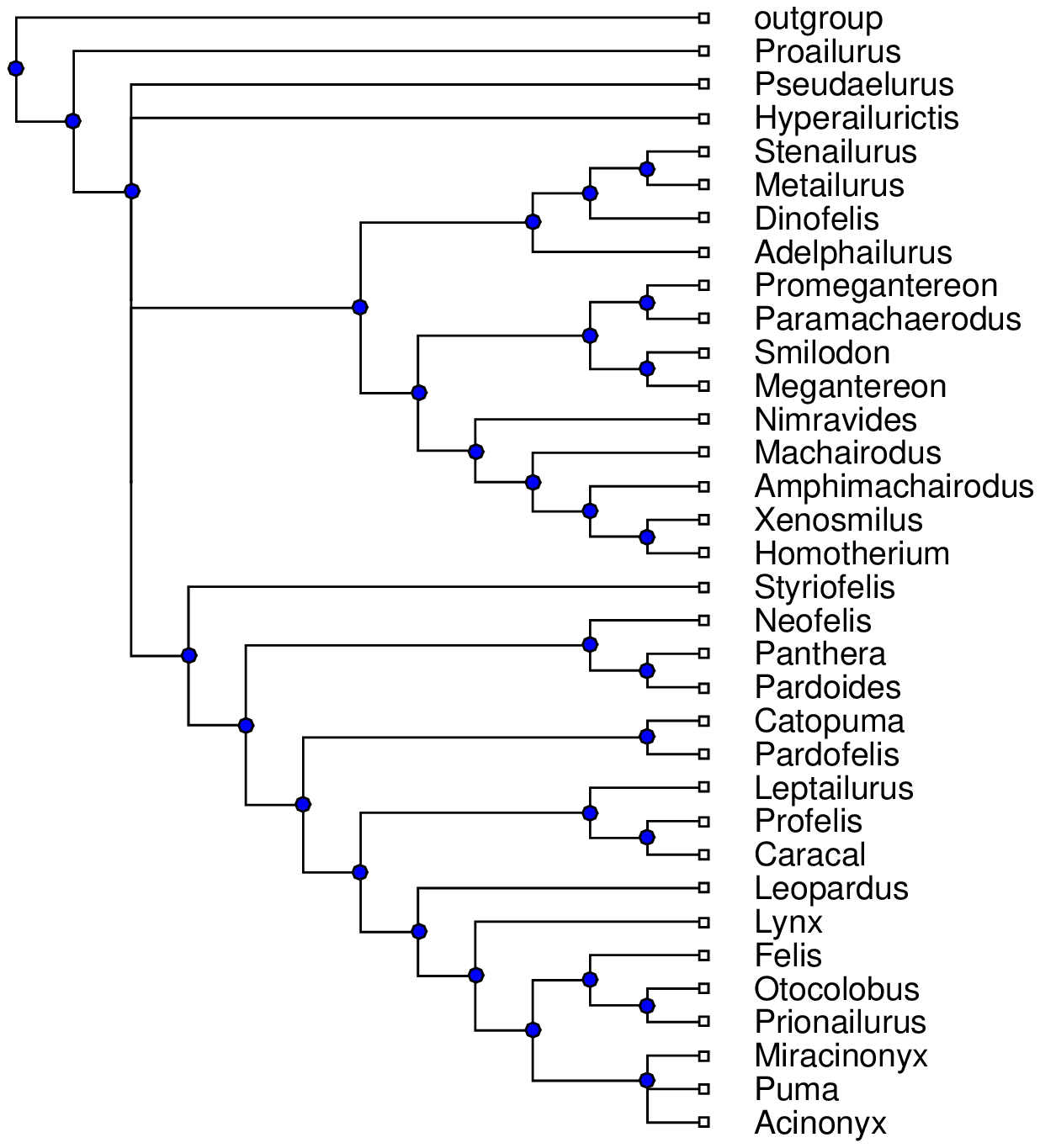}
\end{minipage}%
\begin{minipage}{.5\textwidth}
	\centering
	\includegraphics[width=\textwidth]{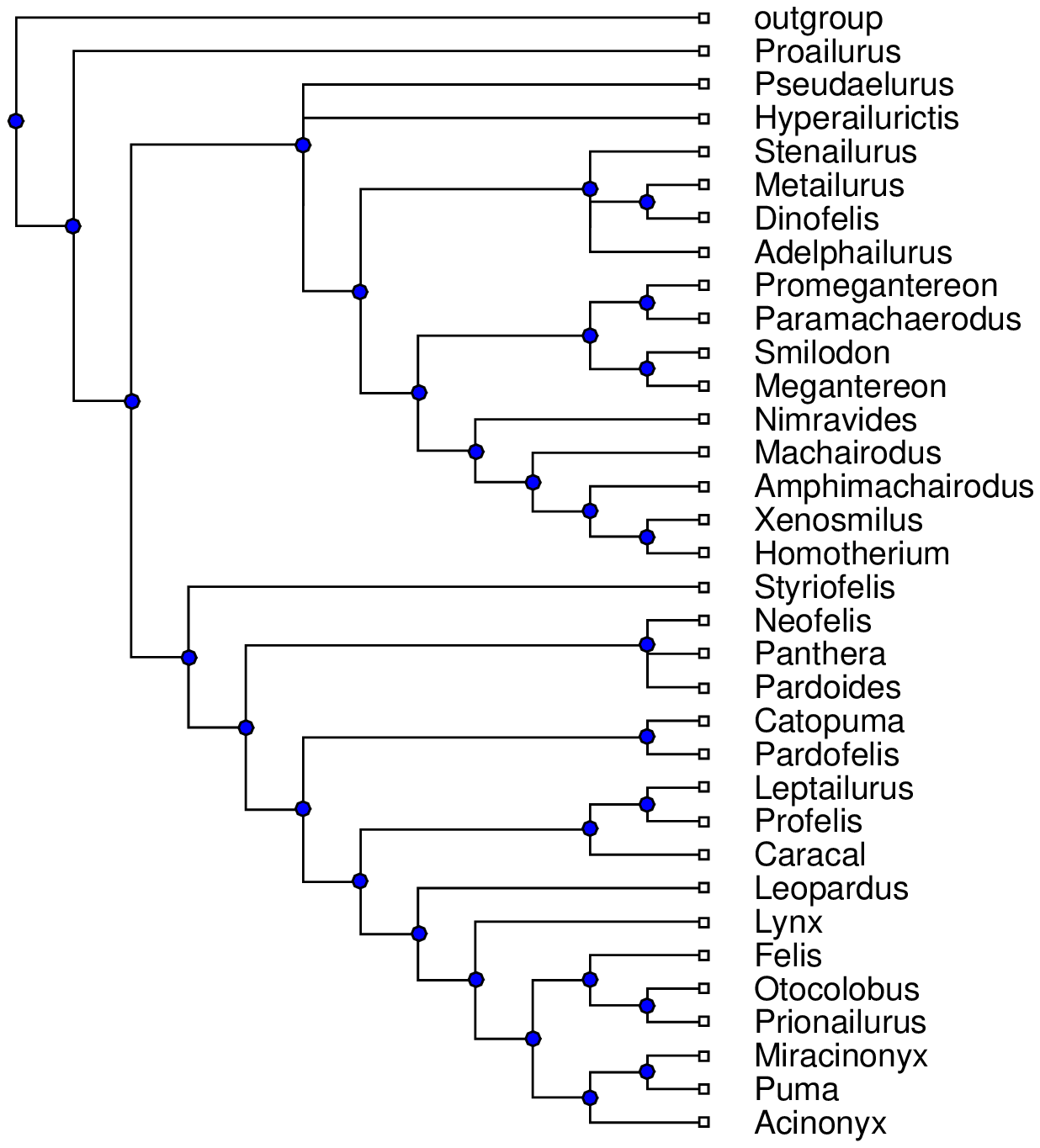}
\end{minipage}
\caption{Left: Best-resolution 50\% majority consensus MRP genus-level
  supertree modified from 
 \protect\cite{saila2011,saila2012} using scheme {\bf All-F-Mol-bb-wgt}; 
Right: The optimal genus-level supertree using projection encoding
and scheme {\bf All-FM-bb-wgt}.
}
\label{fig:supertrees}
\end{figure}

We use \system{clasp} for the computation of all optimal models.
The considered schemes turned out to be unfeasible for the quartet-based
encoding (no optimum was reached by a timeout of 48 hours),
and only results from the projection encoding are included.
It turns out that there exists a {\em unique optimum} for the
projection encoding for both schemes. In the {\bf All-FM-bb-wgt}
scheme, the global optimum was identified in 4~hours and 56~minutes, 
while it was located in 52 minutes for {\bf F-Mol} using 
{\tt --config=trendy} which performed best on these
instances. 
The respective run times are 1.5 hours and 20 minutes using 
parallel \system{clasp} 3.1.2 with 16 threads.

The MRP supertrees in \cite{saila2011,saila2012} are computed using
the full species-level data with the {\em Parsimony Ratchet} method
\cite{nixon1999parsimony}. 
For the resulting shortest trees, 50\% majority consensus trees were
computed and the {\em best supported supertree} according to
\cite{wilkinson2005measuring} out of different runs
(with various MRP settings) originates from scheme {\bf
  All-FM-bb-wgt}, while the {\em best resolved tree} was obtained 
using scheme {\bf All-F-Mol-bb-wgt}. 
Finally, the species-level supertree is collapsed to the genus level.
The optimal supertree for the projection encoding and 
the MRP supertrees from \citeANP{saila2011} described above
(projected to the set of genera considered in our experiments) are
presented in Figure \ref{fig:supertrees} and the online appendix
(Appendices A--C). 

As the true supertree is not known for this real-life dataset, the
goodness of the output tree can only be measured based on how it
reflects the source trees.
To assess the quality of the output trees and to compare them with the 
MRP trees, we considered the number of satisfied quartets of source
trees, the resolution of the supertree, and support values
\cite{wilkinson2005measuring}. Support varies between 
$1$ and~$-1$, indicating good and poor support, respectively, of the
relationships in source trees. 
The results are given in Table \ref{results}, showing that the optimum
of the projection encoding satisfies more quartets of the input data than
the MRP supertrees.

\begin{table}
\begin{minipage}{\textwidth}
\begin{tabular}{lccccc} 
Scheme & Method & Resolution & 
QS\footnote{Number of satisfied quartets from source trees}
& \%QS\footnote{Percentage of satisfied quartets from source trees} 
& V\footnote{Support according to \cite{wilkinson2005measuring}}
\\ \hline\hline
{\bf All-FM-bb-wgt}
& proj & 0.90 & 14 076 
& 0.84 & 0.43 \\
{\bf All-FM-bb-wgt} 
& MRP & 0.85 
& 12 979 
& 0.77 
& 0.45 
\\ 
{\bf All-F-Mol-bb-wgt} 
& MRP & 0.93 
& 13 910 
& 0.83 
& 0.42 
\\ \hline
{\bf F-Mol}
& proj & 1.00    & 4 395 
& 0.86 & 0.25 \\
{\bf F-Mol}            
& MRP  & 1.00    & 4 389 
& 0.86 & 0.27\\ \hline
\end{tabular}
\vspace{-2\baselineskip}
\end{minipage}
\caption{Comparison between the optimal supertree for the projection
  encoding (proj) and the best MRP supertrees.}
\label{results}
\end{table}

Finally, the differences of the objective functions of our two
encodings can be illustrated by computing the supertree of 5 highly
conflicting source trees of 8 species of hammerhead sharks from
\cite{cavalcanti}.  
The optimum for the projection encoding is exactly the same as source
tree (b) in \cite{cavalcanti}, whereas the optimum for quartet
encoding is exactly the same as source tree (a). 
Thus, the two objective functions are not equivalent in the case of
conflicting source trees.

\section{Conclusion}
\label{sect:conc}

In this paper we propose two ASP encodings for phylogenetic supertree 
optimization. The first, solving the maximum quartet consistency
problem, is similar to the encoding in \cite{wu2007quartet} and  
does not perform too well in terms of run time when the size of the
input (source trees and number of taxa therein) grows. The other
novel encoding is based on projections of trees and the
respective optimization problem is formalized as the maximum
projection consistency problem.  
We use real data, namely a collection of phylogenetic trees for the
family of cats (Felidae) and first evaluate the performance of our
encodings by computing genus-specific supertrees.
We then compute a genus-level supertree for the
data and compare our supertree against a recent
supertree computed using MRP approach \cite{saila2011,saila2012}. 
The projection-based encoding performs better than the quartet-based
one and produces a unique optimum for the two cases we consider (with
rogue taxa removed). Obviously, this is not the case in general and in
the case of several optima, consensus and majority consensus 
supertrees can be computed. 
Furthermore, our approach produces supertrees comparable to ones
obtained using MRP method.
For the current projection-based encoding, the problem of optimizing a
species-level supertree using the Felidae data is not feasible as a
single batch. 
Further investigations how to tackle the larger species-level data are
needed. Possible directions are for instance using an incremental
approach and/or parallel search. 

\section{Acknowledgments}

This work has been 
funded by the Academy of Finland, grants 251170
(Finnish Centre of Excellence in Computational Inference Research
COIN), 132995~(LS), 275551~(LS), and 250518~(EO). We thank
Martin Gebser, Ian Corfe, and anonymous reviewers for discussion and
comments that helped to improve the paper.

\bibliography{paper39}

\end{document}